\documentclass{article}
\usepackage{arxiv}

\usepackage[utf8]{inputenc} 
\usepackage[T1]{fontenc}    
\usepackage{hyperref}
\usepackage{url}              
\usepackage{graphicx}
\usepackage{natbib}
\usepackage{doi}
\usepackage{bm}
\usepackage{amsmath}
\usepackage{float}

\usepackage{setspace}
\spacing{1.3}

\title{\textbf{GRFsaw}: \\ A lightweight stochastic microstructure generator}
\date{ \small
    $^1$Institute for Geotechnical Engineering, ETH Zürich, Zürich, Switzerland  \\[1ex]%
    $^2$WSL Institute for Snow and Avalanche Research SLF, Davos, Switzerland  \\[1ex]%
    $^3$Climate Change, Extremes, and Natural Hazards in Alpine Regions Research Center CERC, Davos, Switzerland \\[3ex]%
    \normalsize
    \today
}

\author{ 
\href{https://orcid.org/0000-0002-2016-7233}{\includegraphics[scale=0.06]{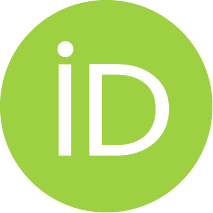}\hspace{1mm}Lars Blatny$^{1,2,3}$}\thanks{Email: \texttt{lars.blatny@slf.ch}} \\
	\And
	\href{https://orcid.org/0000-0001-7515-6809}{\includegraphics[scale=0.06]{orcid.pdf}\hspace{1mm}Henning L\"owe$^{2}$} \\
	\And
	\href{https://orcid.org/0000-0001-8931-752X}{\includegraphics[scale=0.06]{orcid.pdf}\hspace{1mm}Johan Gaume$^{1,2,3}$} \\
}

\renewcommand{\headeright}{}
\renewcommand{\undertitle}{}
\renewcommand{\shorttitle}{GRFsaw: A lightweight stochastic microstructure generator}

\hypersetup{
pdftitle={GRFsaw},
pdfauthor={Lars Blatny},
}

\begin{document}
\maketitle

\begin{abstract}
This article presents \texttt{GRFsaw}, an open-source software for generating two-phase (binary) microstructures with user-defined structural properties. 
Unlike most standard software for microstructure generation, \texttt{GRFsaw} is based on the concept of thresholding Gaussian random fields (GRF). 
It is designed to be used by researchers or engineers in need of a lightweight tool to generate microstructures of various geometries, for example as input to simulations or to other models where such geometries are needed.
This could be simulations of fluid flow through porous media, in predictive models of electromagnetic scattering by materials, or in mechanical loading simulations in order to assess, e.g., the material's elasticity or strength.
\end{abstract}


\vspace{1cm}

\section{Introduction}
Snow, ceramics, concrete, human bone, wood, rocks like sandstone or granite, and metals like aluminum or gold, are all examples of porous materials.
On the smaller scales, these materials have special structural geometries and compositions that are typically called the \emph{microstructure}.
The properties of porous materials may depend on, or in some cases be completely dominated by, their microstructure.
Such properties include the material's ability to scatter electromagnetic radiation~\citep{ding_electromagnetic_2010, tan_uniaxial_2016}, their fluid transport properties~\citep{printsypar_2019}, their elastic modulus~\citep{roberts_computation_2002, soyarslan_3d_2018}, their failure strength~\citep{blatny2021}, their plastic consolidation~\citep{blatny2023} and compaction patterns~\citep{blatny2022}.
While experimental measurements can be made to obtain the microstructural geometry of real materials, e.g., through micro-computed X-ray tomography, this can become an expensive and tedious process, especially if many samples are needed.
For fragile or temperature-sensitive materials like snow, although possible, such X-ray measurements require extra care.
Therefore, artificial computational schemes for generating microstructures are useful as an alternative to experimental measurements.

A popular method for constructing microstructures is the scheme of \citet{yeong1998} where simulated annealing is used to solve an energy minimization problem that approximates given $n$-point correlation functions.
While this is a standard and flexible method, it quickly becomes computationally expensive.
Several (open-source) software for generating microstructures are available, including e.g., \texttt{DREAM.3D}~\citep{groeber2014dream}, and more recently, \texttt{Kanapy}~\citep{Biswas2020} and \texttt{MicroStructPy}~\citep{Hart2020swX}.
The two former are based on Voronoi/Laguerre tessellation of distributed grains with given sizes and shapes, while the latter relies on a collision driven particle dynamics approach.
As an alternative way of obtaining two-phase microstructures, \texttt{GRFsaw} is based on thresholding Gaussian random fields (GRF).
The concept (and by extension the numerical implementation) of this scheme is admirably simple: it amounts to constructing a GRF that leads to prescribed structural properties of the resulting microstructure by appropriately sampling the wavevectors of the GRF.
The use of pattern formation through thresholding GRFs can be traced back to~\citet{cahn_phase_1965}, and was later studied by~\citet{roberts1995}, \citet{grigoriu_random_2003}, \citet{jiang_efficient_2013} and \citet{hyman_stochastic_2014}.
In \texttt{GRFsaw}, the user can a priori choose the porosity, the size and distribution of the microstructural ``grains", as well as the direction and degree of preferred grain elongations (i.e., anisotropy).
As a stochastic generator, many different samples with identical parameters can be produced. This is particularly useful if ensembles of structures are needed for statistical analyses.

\vspace{0.2cm}

\section{Theory}
\subsection{Microstructure generation}
\subsubsection{Constructing the GRF}
Given a Gaussian random field $GRF(\bm{r})$, a binary microstructure $Z(\bm{r})$ is obtained by thresholding $GRF(\bm{r})$, i.e.,
\begin{equation}
    Z(\bm{r}) =
    \begin{cases}
        1,     & \text{if \ } GRF(\bm{r}) > c(\phi) \\
        0,     & \text{otherwise }
    \end{cases}
\label{eq:singlecut}
\end{equation}
where $Z(\bm{r}) = 1$ denotes solid phase and $Z(\bm{r}) = 0$ void phase.
The threshold $c(\phi) = \text{erf}^{-1}(1-2\phi)$ depends on the desired solid volume fraction $\phi = V_\text{solid} / V_\text{total}$.
Here, $GRF(\bm{r})$ is created as a superposition of $N$ (typically $\sim 10^3$) standing waves,
\begin{equation}
    GRF(\bm{r}) = \frac{1}{\sqrt{N}} \sum\limits_{n=1}^{N} \cos(\bm{q}_n \cdot \bm{r} + \varphi_n)
\label{eq:grf}
\end{equation}
where the $\varphi_n$ are i.i.d. random variables in $U[0, 2\pi]$ and $\bm{q}_n = 2 \pi / \lambda_n \hat{\bm{q}}_n$ are wavevectors. How the wavevectors' magnitude $2 \pi / \lambda_n$ and direction $\hat{\bm{q}}_n$ should be sampled is outlined below.

\vspace{0.2cm}

\subsubsection{Sampling the wavevectors}
Let $\langle \lambda \rangle$ denote the average sampled wavelength.  
Relative to the length $L$ of the microstructure, this gives an average number of microstructural elements (or ``grains") per unit length, $\langle m \rangle  = L / \langle \lambda \rangle$.
This provides an intuitive understanding of how the wavevector magnitudes should be sampled.
In particular, probability distributions with mean $\langle m \rangle$ and standard deviation $\Delta m$ are introduced as a way to set the relative ``grain" size and the degree of heterogeneity within a structure.
\texttt{GRFsaw} offers the choice between two probability distributions: a normal distribution and the gamma distribution of~\citet{ding_electromagnetic_2010}.
The latter distribution provides an analytic expression of the resulting two-point correlation function, while the former allows for speedier sampling.

\newpage

Isotropic structures require wavevector directions $\hat{\bm{q}}_n$ uniformly sampled from the unit sphere.
However, in order to produce structures with anisotropy (i.e., with a preferred orientation of the microstructural elements), the sampling can be restricted to a subset of the unit sphere as suggested by \citet{tan_uniaxial_2016}.
In \texttt{GRFsaw}, an anisotropy parameter \mbox{$a \in (0,1]$} is introduced, where $a=1$ represents uniform sampling from the full unit sphere and thus producing isotropic structures.
While the software is designed such that the user can choose from structures with either horizontally or vertically preferred directions, in principle any preferential direction can be devised by cleverly sampling~$\hat{\bm{q}}_n$ from the unit sphere.

\subsubsection{Single-cut vs double-cut structures}
The generated structures features thick grains connected by neck regions of smaller thickness.
Here, such microstructures following Equation~\eqref{eq:singlecut} are called \emph{single-cut structures} as only a single threshold is used.
However, it is also possible to generate \emph{double-cut structures},
\begin{equation}
    Z(\bm{r}) =
    \begin{cases}
        1,     & \text{if \ } c_l(\phi) < GRF(\bm{r}) < c_u(\phi) \\
        0,     & \text{otherwise }
    \end{cases}
\label{eq:doublecut}
\end{equation}
with the upper and lower thresholds $c_u(\phi) = c_l(-\phi) = \text{erf}^{-1}(\phi)$.

\begin{figure}[H]
	\centerline{\includegraphics[width=0.99\textwidth]{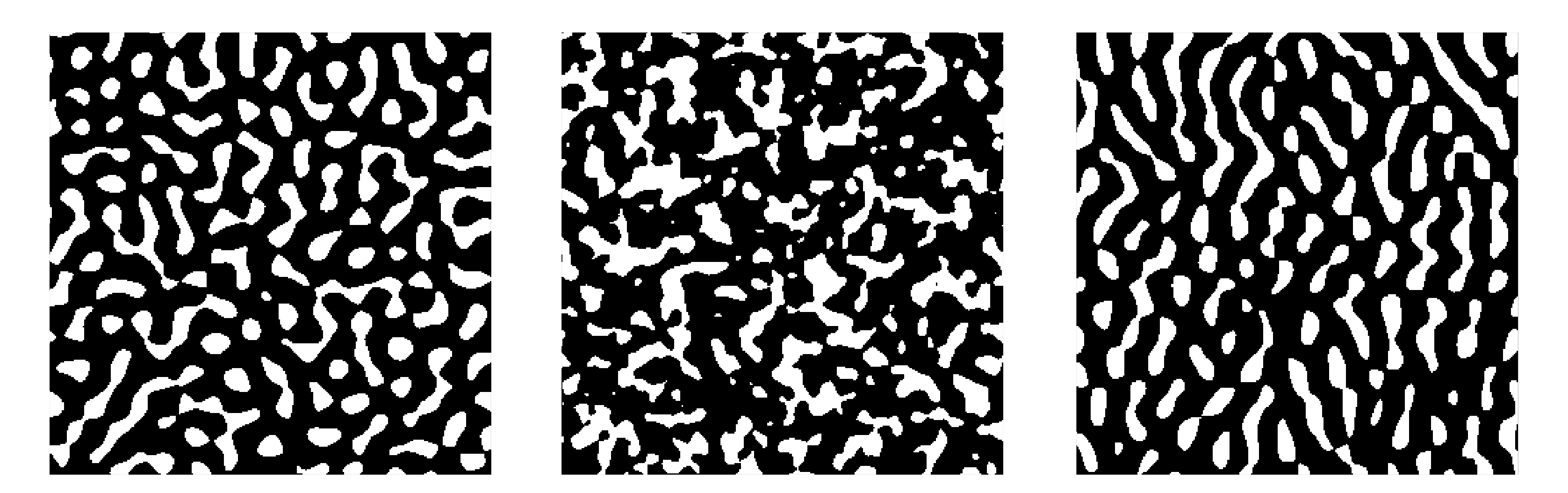}}
	\caption{2D single-cut structures with $\phi = 0.7$ using the gamma distribution and~$\langle m \rangle=13$. From left to right: (a)~an isotropic and fairly homogeneous sample with~$\Delta m = 1.8$, (b) an isotropic and rather heterogeneous sample with $\Delta m = 6.0$ and (c) an anisotropic sample with a preferred vertical direction with $a=0.6$ and $\Delta m = 1.8$.} 
	\label{fig:2d_sc}
\end{figure}

\begin{figure}[H]
	\centerline{\includegraphics[width=0.99\textwidth]{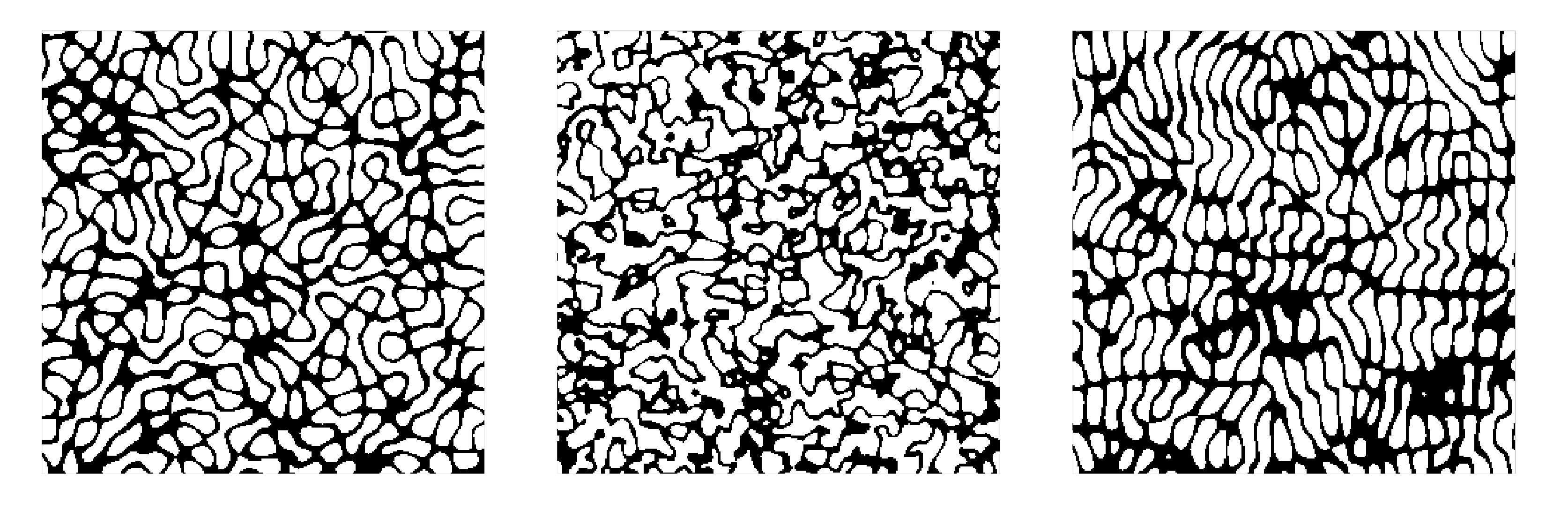}}
	\caption{2D double-cut structures with $\phi = 0.4$ using the gamma distribution and~$\langle m \rangle=13$. From left to right: (a)~an isotropic and fairly homogeneous sample with~$\Delta m = 1.8$, (b) an isotropic and rather heterogeneous sample with~$\Delta m = 6.0$ and (c) an anisotropic sample with a preferred vertical direction with $a=0.6$ and $\Delta m = 1.8$.} 
	\label{fig:2d_dc}
\end{figure}

\begin{figure}[H]
	\centerline{\includegraphics[width=0.99\textwidth]{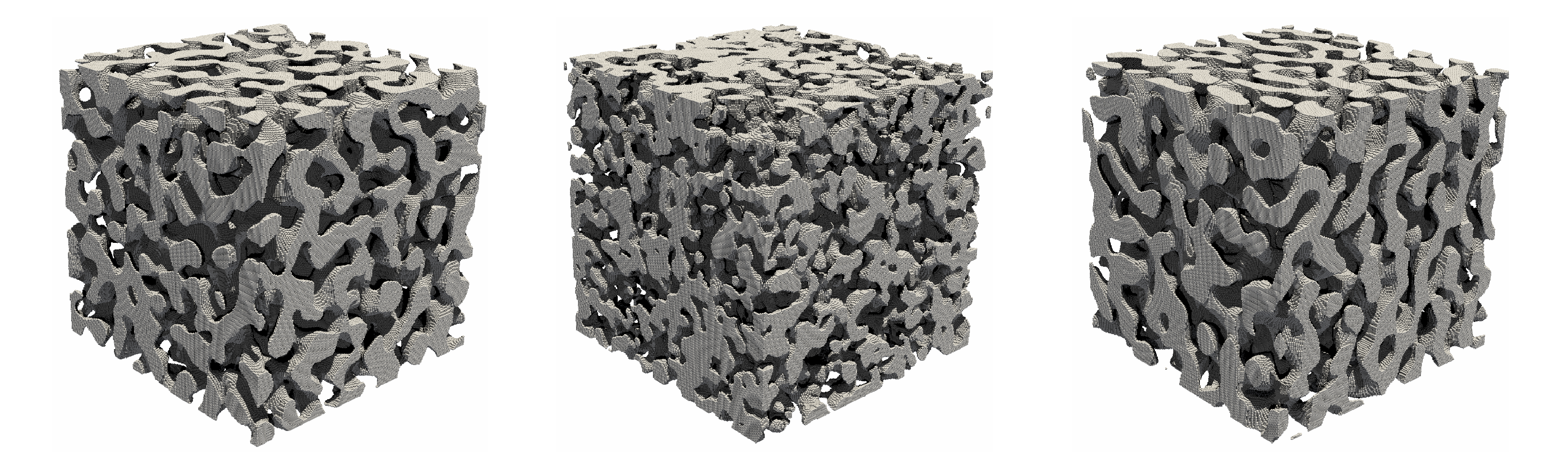}}
	\caption{3D single-cut structures with $\phi = 0.4$ using the normal distribution and $\langle m \rangle=9$. From left to right: (a)~an isotropic and fairly homogeneous sample with~$\Delta m = 1.3$, (b) an isotropic and rather heterogeneous sample with~$\Delta m = 6.0$ and (c) an anisotropic sample with a preferred vertical direction with $a=0.6$ and $\Delta m = 1.3$.} 
	\label{fig:3d}
\end{figure}

The double-cut structures display thin snaky elements with roughly the same thickness.
These elements intersect at denser areas.
For high solid volume fractions, both single- and double-cut isotropic structures resemble a ``Swiss cheese", i.e., a solid perforated with rounded holes.
Assuming sufficient spatial discretization, double-cut structures are guaranteed to percolate.
However, single-cut structures have a percolation threshold at approximately $\phi \approx 0.50$ in 2D and $\phi \approx 0.15$ in 3D.
Figures~\ref{fig:2d_sc} and~\ref{fig:2d_dc} show various 2D single-cut and double-cut microstructures, respectively, with different degree of heterogeneity and anisotropy.
In addition, Figure~\ref{fig:3d} shows 3D single-cut microstructures.

\renewcommand{\arraystretch}{1.3}
\begin{table}[t]
\centering
  \begin{tabular}{lccl}
      \hline
      \textbf{Parameter}    & \textbf{Symbol} & \textbf{Range}   & \textbf{Interpretation}   \\
      \hline 
       Size           & $\langle m \rangle$ & $>0$     & Mean number of grains per horizontal length \\ 
       Heterogeneity  & $\Delta m$          & $>0$     & Standard deviation of number of grains per length \\
       Anisotropy     & $a$                 & $\in (0,1]$  & A grain elongation ratio along preferred direction \\
       Solid fraction      & $\phi$              & $\in(0,1)$  & Solid phase volume relative to total volume  \\
       \hline  \\ 
  \end{tabular}
  \caption{\vspace{10mm} Main microstructural parameters for both single- and double-cut structures.}
  \label{tab:params}
\end{table}
\renewcommand{\arraystretch}{1}

Table~\ref{tab:params} lists the microstructural input parameters which are to be specified by the user. In addition to these parameters, the user must also specify whether the desired structure is single- or double-cut, the degree of spatial discretization (i.e, the number of grid point on a regular grid),  and the preferred direction (horizontal or vertical) of grain elongation in the case of anisotropy, i.e., when~$a<1$. 

\newpage

\subsection{Two-point correlation function}
To compute the two-point correlation function, 
\begin{equation}
    R(\bm{r}) = \frac{1}{V_\text{total}} \int Z(\bm{r'}) Z(\bm{r'} + \bm{r}) d^3 \bm{r'} \text{,}
\end{equation}
\texttt{GRFsaw} provides an efficient calculation by relying on the Wiener-Khinchin theorem, using fast Fourier transforms (FFT) of the binary microstructure.
In the specific case of isotropic single-cut structures based on the gamma distribution function of \citet{ding_electromagnetic_2010} the two-point correlation function can be computed analytically. 

The angular-averaged two-point correlation function is given by
\begin{equation}
    g(r) = \frac{1}{4\pi} \int_0^{2\pi} \int_0^\pi  R(r, \theta, \varphi) \sin(\theta) d\theta d\varphi \text{,}
\end{equation}
and the normalized angular-averaged two-point correlation function is simply 
\begin{equation}
    g_\text{norm}(r) = \frac{g(r) - \phi^2}{\phi - \phi^2} \text{.}
\end{equation}

\begin{figure}[htb]
	\centerline{\includegraphics[width=0.75\textwidth]{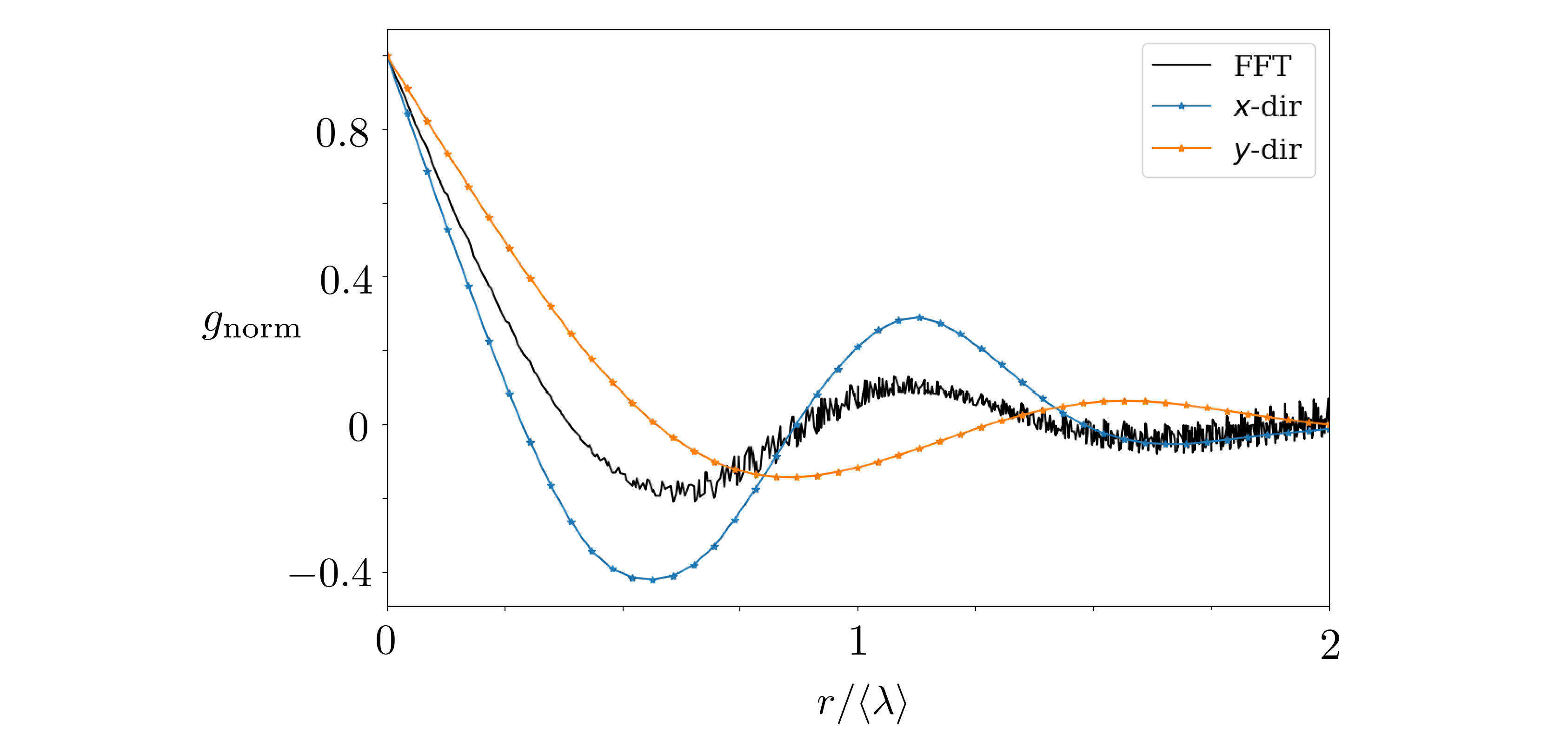}}
	\caption{Normalized two-point correlation function $g_\text{norm}(r)$ of the anisotropic sample in Figure~\ref{fig:2d_sc} (rightmost) Here, both the angular averaged correlation computed with FFT and the 1D-correlations in the $x$- and $y$-directions are computed, thus highlighting the anisotropic geometry.} 
	\label{fig:auto}
\end{figure}

If desired, the structural parameters can be tuned to approximate a normalized angular-averaged two-point correlation.
In addition to the angular averaged two-point correlation, a scheme to compute the one-dimensional two-point correlation
in the $x$-, $y$- and $z$-direction is included, thus allowing for a detailed assessment of anisotropy.
Figure~\ref{fig:auto} shows the computed two-point correlation of the anisotropic structure visualized in Figure~\ref{fig:2d_sc}.

\subsection{Specific surface area and tortuosity}
For many porous materials, the \emph{specific surface area} ($SSA$) is an important microstructural indicator.
In \texttt{GRFsaw}, $SSA$ is defined as the ratio of surface area to solid volume and can be related to the initial slope of the angular-averaged two-point correlation function through
\begin{equation}
    SSA = -\frac{4}{\phi} \lim_{r \to 0} \frac{\partial}{\partial r} g(r)
\label{eq:SSA}
\end{equation}
as suggested by \citet{berryman_relationship_1987} as a generalization of the work by \cite{Debye1957}.

Another important microstructural property is \emph{tortuosity}, which gives the ratio between the (shortest) flow path distance through the material relative to its side length.
Relying on the so-called burning method by \citet{herrmann1984}, \texttt{GRFsaw} allows for the calculation of the shortest path length through a phase (solid phase or void phase) of a structure.

In certain applications, microstructures with grains unconnected to (one of) the percolating cluster(s) may be undesirable.
As a remedy, the burning method is also employed by \texttt{GRFsaw} to remove parts of the structure that do not belong to the percolating cluster.

\section{Concluding remarks}
This short article has presented \texttt{GRFsaw}, a simple and lightweight open-source software for generating microstructural geometries based on the principle of thresholding Gaussian random fields. 
It allows its users to create microstructures with pre-defined 1) solid volume fraction, 2) heterogeneity through a ``grain size" distribution and 3) anisotropy as a preferred direction of the microstructural elements.
These properties can be related to the microstructure's two-point correlation and surface area. 
Post-processing tools in \texttt{GRFsaw} allow for the computation of the two-point correlation function and specific surface area, as well as the microstructure's tortuosity.

\section{Code availability and documentation}
\texttt{GRFsaw} is available under the MIT license on GitHub (\url{https://github.com/larsblatny/GRFsaw/}) where a detailed documentation can also be found.

\section{Acknowledgments}
L.B. and J.G. acknowledge support from the Swiss National Science Foundation (grant number PCEFP2\_181227) while employed at EPFL.

\bibliographystyle{unsrtnat}
\bibliography{bibfile}

\end{document}